# Orbital-Selective Mottness in $K_xFe_{2-y}Se_2$ Superconductors Revealed by Pump-Probe Spectroscopy

Wei Li,[1] Chunfeng Zhang,[1,]* Shenghua Liu,[1] Xiaxin Ding,[1] Xuewei Wu,[1] Xiaoyong Wang,[1] Hai-Hu Wen,[1] and Min Xiao[1,2,]*

[1]National Laboratory of Solid State Microstructures and Department of Physics, Nanjing University, Nanjing 210093, China

[2] Department of Physics, University of Arkansas, Fayetteville, Arkansas 72701, USA

**Abstract**

**We report transient optical signatures of the orbital-selective Mottness in superconducting $K_xFe_{2-y}Se_2$ crystals by using dual-color pump-probe spectroscopy. Besides multi-exponential decay recovery dynamics of photo-induced quasiparticles, a damped oscillatory component due to coherent acoustic phonons emerges when the superconducting phase is suppressed by increasing the temperature or excitation power. The oscillatory component diminishes with significant enhancement of a slow decay component upon raising temperature to 150-160 K. These results are in consistence with the picture of orbital-selective Mott phase transition, indicating a vital role played by electron correlation in the iron-based superconductors.**

It is a focus issue whether the iron-based superconductors (FeSCs) are in close proximity to Mott behavior driven by electron correlation [1-6]. So far, superconductivities in ferropnictides have been generally understood within the Fermi surface nesting scheme with weak and moderate correlations [7-10]. The recently discovered alkaline iron selenide superconductors [11-14], with their transition temperatures ($T_C$) above 30 K have stimulated an interest to reevaluate the role played by electron correlation [1-6, 15, 16]. Their Fermi-surface structures, as revealed in the multiple angle-resolved- photoemission-spectroscopy (ARPES) experiments [17-20], are different from the common $s^{\pm}$-pairing symmetry. The ordered magnetic moment in $K_xFe_{2-y}Se_2$ (KFS) was found to be the largest among FeSCs [21]. The parent phase of KFS superconductor has been found to be on the border of an insulating and magnetically-ordered state despite the fact that an exact structure of parent phase is still under debate [3, 22-24]. These properties cannot be fully explained by the nesting mechanism. Instead, they are more closer to the behavior of doped Mott insulators like the cuprates [3-5, 25, 26].

Despite a metal-insulator crossover in transport curves has been measured in the superconductor family of alkaline iron selenides [11-14], it is still quite challenging to directly indentify the Mottness by experiments owing to the coexistence of insulating and superconducting/metallic domains in the KFS superconducting samples [3, 24, 27, 28]. The nature of multiple bands crossing the Fermi surface [17-20, 29] makes this issue even more complicated. Mott-like physics has been observed on the insulating KFS samples, which, however, is unlikely to give rise to the superconducting phase[3]. Very recently, Yi *et al.* suggested a scenario with orbital-selective Mott phase to explain the ARPES data on superconducting KFS samples [4], which is in consistence with the prediction of a multiband theory assuming strong on-site Coulomb interactions [5]. Nevertheless, such Mott behavior is not a prerequisite for understanding the resistivity hump ($T_H$ in the range of 50-250 K) observed in the transport curves, which apparently can also be viewed as a result of parallelly-aligned resistors [30, 31]. Femtosecond pump-probe spectroscopy can provide new opportunities to investigate this issue by providing additional dynamic information about quasiparticles (QPs) [32-38], phase transition [39], competing orders [40], and collective modes [41-43]. In particular, these time-resolved optical techniques can monitor the microstructures isolated from each other, adding vital information which is unattainable by electrical characterizations.

In this Letter, we report a femtosecond spectroscopic study on superconducting KFS single crystals that reveals ultrafast optical signatures of the orbital-selective Mottness. Entangled with a biexponential recovery dynamics of photo-excited QPs, a damped oscillatory component emerges when the superconducting phase is suppressed by increasing the temperature or excitation power. Such oscillatory signal is probably induced by coherent acoustic phonons related to competing orders. Upon raising temperature to 160 K, the oscillatory component diminishes together with a dramatic increase of a slow exponential component in the recovery trace of QP dynamics. Such abnormal temperature-dependent behaviors are signatures of the orbital-selective Mott physics with gap opening at the 3d xy bands in KFS superconductors [4, 5]. The confirmation of this orbital-selective Mottness indicates a vital role played by the electron correlation for superconductivity in KFS crystals.

The samples of superconducting KFS crystals, with superconducting transition at $T_C \approx 32K$ and a resistivity hump at $T_H \approx 180K$ (Fig. 1(a)), were prepared by a rapid quenching process as reported previously [14, 24]. To avoid the heating accumulation induced by the high pulse energy required in the study, we performed dual-color pump-probe experiments with low-repetition (1 kHz) femtosecond pulses (~ 100 fs, Libra, Coherent) [44]. The sample was excited with photon energy of 3.1 eV, and the QP dynamics was monitored by measuring the transient reflectivity changes ($\Delta R / R$) with the probe photon energy of 1.55 eV. To exclude the aging effect on the cleaved surface of the samples, we loaded the samples into an optical cryostat (MicroCryostatHe, Oxford) with great care to avoid exposing the samples to ambient environment. We performed the temperature cycle test, and found reproducible experimental data in the first 48 hours after sample loading. More experimental details are available in the Supplementary Materials. In general, the magnitude of $\Delta R/R$ signal is proportional to the density of photo-induced QPs [45]. In superconductors, the QPs are excited through the avalanche multiplication due to the electron-electron collisions [45], so that one incident photon creates a large number of QPs. The number of QPs created per absorbed photon is proportional to the ratio between the excitation photon energy and the gap value near the Fermi level [45]. Since the insulator $K_2Fe_4Se_5$ phase has a large gap (500 meV) [3, 23], the contribution from the insulating domains is negligible.

The validity of the measurements has been confirmed by monitoring the QP dynamics of the KFS superconductors in the weak perturbation regime (Fig. S1, Supplementary Materials). The pump photon energy is sufficiently large to excite the electronic transitions, causing a swift decrease of $\Delta R/R$. Under weak perturbation, it is generally accepted that the recovery of $\Delta R/R$ in superconductors can be explained by the Rothwarf-Taylor model [40, 46, 47]. As discussed in the Supplementary Materials, the QP dynamics in KFS superconductors, as characterized by the signal magnitudes and the decay lifetimes, can also be well understood by this model. By fitting the temperature-dependent magnitude with the model, the gap value is evaluated to be ~ 12 meV, which is in good agreement with the values measured by ARPES experiments [19].

The time-resolved traces of $\Delta R/R$ are significantly dependent on the incident fluence (Fig. S2, Supplementary Materials). In superconducting phase, an oscillatory component emerges when the fluence increases (Fig. 1(b)). We can analyze the decay curves with a phenomenological function of the form [42, 43]

$$\Delta R(t)/R = A_1 e^{-t/\tau_1} + A_2 e^{-t/\tau_2} + A_3 + A_O e^{-t/\tau_O} \cos[2\pi f(t+t_0)] \ , \qquad (1)$$

where $A_1$ and $A_2$ are the magnitudes of two exponential decay components with lifetimes $\tau_1$ (< 10 ps) and $\tau_2$ (< 100 ps) contributed by the excited electrons and phonons, respectively [42, 43]; the constant $A_3$ represents a slow decay component with lifetime longer than the time window. The last item shows a damped oscillatory component with amplitude $A_O$, damping lifetime $\tau_O$, oscillatory frequency $f$, and the initial phase of $2\pi f t_0$. At $T = 5K$, the decay is dominated by the fast exponential component with an excitation-fluence-dependent lifetime. The fluent-dependent $A_1$ and $A_O$ are shown in Fig. 1(c). With increasing incident fluence, the exponential component rises gradually; however, the oscillatory amplitude shows a threshold behavior which becomes detectable when the excitation fluence reaches a threshold value of $I_{th} \sim 40$ μJ/cm$^2$. The decay rate ($1/\tau_1$) is also strongly dependent on the incident fluence, indicating an important role played by electron-electron interactions for the recombination of photo-excited QPs [34]. The fluence dependence of the decay rate ($1/\tau_1$) also exhibits a slope change at $I_{th}$. These significant changes on the QP dynamics at $I_{th}$ observed here can be assigned to the vaporization of superconducting phase, since $I_{th}$ is close to or even higher than the fluence required for vaporizing the superconducting phase in cuprates [39, 47] and other FeSCs [36, 37]. In other words, the oscillatory component emerges when the superconductivity is suppressed.

To check this, we then study the temperature dependence at a high incident fluence of ~ 200 μJ/cm$^2$. The magnitude of $\Delta R/R$ is plotted as functions of delay time and temperature in Fig. 2(a). The signal has an initial negative value close to zero time delay (t ~ 0.2 ps) and a positive value at the delay time of t ~ 4.5 ps induced by the oscillatory component. These two features show quite different temperature dependences (Figs. 2(b) & 2(c)). The initial negative value becomes less dominant when the temperature increases (Fig. 2(b)). Since the fluence is large enough to destroy the superconducting phase, the temperature-dependent magnitude can be analyzed with the Mattis-Bardeen formula in the form of [36, 37]

$$\frac{\Delta R(T)}{R} \propto (\frac{\Delta(T)}{\hbar\omega})^2 \log(\frac{3.3\hbar\omega}{\Delta(T)}), \qquad (2)$$

where $\hbar\omega$ is the probe-photon energy. The equation can reproduce the experimental data quite well with the superconducting gap value of 12 meV (Fig. 2(b)). In contrast, the signal amplitude at t ~ 4.5 ps increases dramatically upon raising temperature (Fig. 2(c)), which is a signature for the increase of oscillatory component. The amplitude of the oscillatory component ($A_O$), obtained by fitting the data with Eq.1, increases dramatically when temperature crosses over $T_C$ (Fig. 2(c)). Such difference also provides an evidence for the competition between the order related to the oscillatory component and the superconducting phase.

Due to the lack of either static magnetic order or excitonic resonance in the superconducting phase, such low frequency oscillation observed in superconductors has generally been recognized as a result of coherent acoustic phonon vibration due to stimulated Brillouin scattering [35, 42, 43]. In principle, this phonon mode can come from either the superconducting domains or the insulating domains in the inhomogeneous KFS superconductors. If contributed from the insulating domains, a gradual increase from zero should be expected with the increasing excitation fluence, which is in contrast to the threshold behavior observed here (Fig. 1(c)). The fluence- and temperature-dependences of the oscillatory amplitude indicate that the contribution from insulator phase can be ruled out here. The abrupt change of signal at $T_C$ (Fig. 2(c)) is a distinct signature of correlated interplay between superconducting phase and the oscillatory component. Thus, we can safely assign the oscillatory component to be due to the contribution from superconducting domains, which is tightly associated with the competing orders since it is activated when the superconductivity is suppressed. Moreover, we observed that the oscillatory frequency, i.e., the phonon energy (Fig. 2(d)), gradually decreases with increasing temperature. Such lattice hardening in the superconducting phase is consistent with ultrasound experiments on superconducting iron arsenides [48], confirming its origin of the superconducting domains.

We now shift our attention to the case of higher temperature. The experimental data obtained at 80 K and 240 K are compared in Fig. 3(a). We carefully analyze the experimental results with Eq.1 and highlight the exponential decay components. There are two major differences between results at the two temperatures: (1) the oscillatory component becomes negligible at higher temperature, and (2) The fast exponential component gets less dominant but a slow decay component with a lifetime of ~ 80 ps becomes pronounced at high temperature. To clarify such temperature-dependent behaviors, the time-resolved traces of $\Delta R/R$ recorded at different temperatures are compared with a normalized scale (Fig. 3(b)). Upon raising temperature, the amplitude of the fast component decreases, while the ratio of $A_2/A_1$ increases and becomes unchanged at a temperature slightly below $T_H$ (Figs. 3(c) & 3(d)). Such temperature-dependent QP dynamics with significant increase of $A_2/A_1$ might be associated to the Mott behavior with gap opening [49].

At the same temperature range, the magnitude of oscillatory component shows a critical point behavior. To quantify the temperature-dependent behaviors, we subtract the exponential decay part and plot the oscillatory component in Fig. 4(a). The amplitude and frequency as analyzed by Fourier transformation are shown in Figs. 4(b) & 4(c), respectively. The oscillation amplitude becomes less important and eventually vanishes at higher temperature (Fig. 4(b)). In principle, depletion of phonon response may be induced by a structure change, which, however, is unlikely here. On one hand, in this temperature range, no structure transition has ever been reported in KFS superconductors [23, 50]; On the other hand, structure transitions should always accompany with a large modification of phonon energy [42], which has not been observed here (Fig. 4(c)). The resistivity change within this temperature range has been explained as a result of parallelly-aligned resistors consisting of insulating and superconducting/metallic domains [30, 31]. However, such a picture cannot fully explain the abrupt change of optical response observed here. The pump-probe measurements monitor the separated domains simultaneously, and the ultrafast dynamical behaviors in either insulator or metal should change gradually with temperature. As reported by Kim et al. [51], the generated coherent phonons can reflect the Mott transition in strongly-correlated electronic systems.

Let's try to understand the observed phenomena using the Mott picture in KFS superconductors. The ARPES experiment done by Yi *et al.* has shown orbital-selective Mott physics in the KFS superconductors [4, 5]. Similar to many other FeSCs, multiple orbitals contribute spectral weights near Fermi surface in FKS samples [17-20]. When raising temperature, the spectral weight near Fermi surface for 3d xy orbitals diminishes while the other orbitals remain metallic. This scenario of orbital-selective Mottness can naturally explain our results described above. For KFS superconductors, all the 3d bands are metallic at low temperature above $T_C$. The photo-excited QPs in metals reach equilibrium with energy dissipation to the lattice due to electron-phonon interaction, which induces a coherent lattice vibration, manifesting itself as an oscillation in the time-resolved trace of $\Delta R/R$. With multiple orbitals crossing the Fermi surface, QP dynamics in FeSCs is band dependent, which has been investigated in iron arsenides [34]. The photo-induced effects revealed in the time-resolved traces are related to the opening of a gap (or pseudogap) in the density of states close to the Fermi energy [45]. When temperature increases, the band renormalization opens a gap larger than 50 meV at the 3d xy bands [4]. Other processes, like interband transition and intervalley scattering, should account for the recombination of QPs [52]. Less energy will drive the lattice vibrations, leading to the reduction of oscillatory amplitude. The gap opening at the xy bands will slow down the recombination of QPs trapped at the gap states, leading to the emergence of the slow exponential component. The fast decay component, mainly contributed from the metallic bands and hot QPs, becomes less important at higher temperature.

From the above discussions, the intriguing temperature-dependent behaviors of QP dynamics [49] and the coherent phonon dynamics [51] can be plausibly attributed to the orbital-selective Mottness in KFS superconductors. Similar results obtained from different sample batches reveal the generality of the above Mott behaviors in KFS superconductors. Physically, the Mott phase, induced by the strong correlation effects, can dramatically modify the transport properties. The coexistence of Mott-localized electrons at 3d xy orbitals and itinerant electrons at the other 3d orbitals involved in an orbital-selective Mott transition should be reflected in transport curves. This does not contradict with the model of parallelly-aligned resistors, as multiple bands crossing the Fermi level, the orbital selective Mott phase is still metallic. To explain the insulating behavior above $T_H$, the parallel resistor model including the insulating domains is still indispensable; however, the Mottness should be considered for temperature dependence of resistivity in metallic domains.

In summary, we have observed signatures for a transition to orbital-selective Mott phase in superconducting KFS single crystals by employing dual-color pump-probe spectroscopy. When the superconducting phase is suppressed by increasing the temperature or increasing the excitation fluence, the coherent acoustic phonon signal, characterized by a damped oscillatory component, appears in the time-resolved traces of $\Delta R/R$. Upon increasing the temperature to 160-180 K, the oscillatory signal diminishes while a slow decay component becomes pronounced. These experimental results can be naturally explained by the orbital-selective Mott physics in KFS superconductors with band normalization at 3d xy bands where the opening of a gap slows down the QP recombination. Our results provide valuable information about the competing orders and confirm the proximity to Mott behaviors induced by the electron correlation for superconductivity in KFS superconductors.

This work is supported by the National Basic Research Program of China (2012CB921801 and 2013CB932903, MOST), the National Science Foundation of China (9123310, 61108001, 11227406 and 11021403), and the Program of International S&T Cooperation (2011DFA01400, MOST). C.Z. acknowledges financial support from the New Century Excellent Talents program (NCET-09-0467), Fundamental Research Funds for the Central Universities, and the Priority Academic Program Development of Jiangsu Higher Education Institutions (PAPD). C.Z. acknowledges Dr. Dawei Shen, Dr. Huan Yan and Prof. Jianxin Li for stimulated discussions.

# References and Notes

* Corresponding authors: cfzhang@nju.edu.cn (CZ), and mxiao@uark.edu (MX)

[1] J.-X. Zhu, R. Yu, H. Wang, L. L. Zhao, M. D. Jones, J. Dai, E. Abrahams, E. Morosan, M. Fang, and Q. Si, Phys. Rev. Lett. **104**, 216405(2010).

[2] R. Yu, J.-X. Zhu, and Q. Si, Phys. Rev. Lett. **106**, 186401(2011).

[3] F. Chen, M. Xu, Q. Q. Ge, Y. Zhang, Z. R. Ye, L. X. Yang, J. Jiang, B. P. Xie, R. C. Che, M. Zhang, A. F. Wang, X. H. Chen, D. W. Shen, J. P. Hu, and D. L. Feng, Phys. Rev. X **1**, 021020(2011).

[4] M. Yi, D. H. Lu, R. Yu, S. C. Riggs, J. H. Chu, B. Lv, Z. K. Liu, M. Lu, Y. T. Cui, M. Hashimoto, S. K. Mo, Z. Hussain, C. W. Chu, I. R. Fisher, Q. Si, and Z. X. Shen, Phys. Rev. Lett. **110**, 067003(2013).

[5] R. Yu, and Q. Si, Phys. Rev. Lett. **110**, 146402(2013).

[6] H. Lei, M. Abeykoon, E. S. Bozin, K. Wang, J. B. Warren, and C. Petrovic, Phys. Rev. Lett. **107**, 137002(2011).

[7] M. M. Qazilbash, J. J. Hamlin, R. E. Baumbach, L. Zhang, D. J. Singh, M. B. Maple, and D. N. Basov, Nat. Phys. **5**, 647(2009).

[8] W. L. Yang, A. P. Sorini, C. C. Chen, B. Moritz, W. S. Lee, F. Vernay, P. Olalde-Velasco, J. D. Denlinger, B. Delley, J. H. Chu, J. G. Analytis, I. R. Fisher, Z. A. Ren, J. Yang, W. Lu, Z. X. Zhao, J. van den Brink, Z. Hussain, Z. X. Shen, and T. P. Devereaux, Phys. Rev. B **80**, 014508(2009).

[9] J. Dong, H. J. Zhang, G. Xu, Z. Li, G. Li, W. Z. Hu, D. Wu, G. F. Chen, X. Dai, J. L. Luo, Z. Fang, and N. L. Wang, Europhys. Lett. **83**, 27006(2008).

[10] S. Graser, T. A. Maier, P. J. Hirschfeld, and D. J. Scalapino, New J. Phys. **11**, 025016(2009).

[11] J. Guo, S. Jin, G. Wang, S. Wang, K. Zhu, T. Zhou, M. He, and X. Chen, Phys. Rev. B **82**, 180520(2010).

[12] A. Krzton-Maziopa, Z. Shermadini, E. Pomjakushina, V. Pomjakushin, M. Bendele, A. Amato, R. Khasanov, H. Luetkens, and K. Conder, J. Phys. Condens. Matt. **23**, 052203(2011).

[13] H.-D. Wang, C.-H. Dong, Z.-J. Li, Q.-H. Mao, S.-S. Zhu, C.-M. Feng, H. Q. Yuan, and M.-H. Fang, Europhys. Lett. **93**, 47004(2011).

[14] H.-H. Wen, Rep. Prog. Phys. **75**, 112501(2012).

[15] J. Hu, and N. Hao, Phys. Rev. X **2**, 021009(2012).

[16] Y. Zhou, D.-H. Xu, F.-C. Zhang, and W.-Q. Chen, Europhys. Lett. **95**, 17003(2011).

[17] D. Mou, S. Liu, X. Jia, J. He, Y. Peng, L. Zhao, L. Yu, G. Liu, S. He, X. Dong, J. Zhang, H. Wang, C. Dong, M. Fang, X. Wang, Q. Peng, Z. Wang, S. Zhang, F. Yang, Z. Xu, C. Chen, and X. J. Zhou, Phys. Rev. Lett. **106**, 107001(2011).

[18] T. Qian, X. P. Wang, W. C. Jin, P. Zhang, P. Richard, G. Xu, X. Dai, Z. Fang, J. G. Guo, X. L. Chen, and H. Ding, Phys. Rev. Lett. **106**, 187001(2011).

[19] Y. Zhang, L. X. Yang, M. Xu, Z. R. Ye, F. Chen, C. He, H. C. Xu, J. Jiang, B. P. Xie, J. J. Ying, X. F. Wang, X. H. Chen, J. P. Hu, M. Matsunami, S. Kimura, and D. L. Feng, Nat. Mater. **10**, 273(2011).

[20] Z. H. Liu, P. Richard, N. Xu, G. Xu, Y. Li, X. C. Fang, L. L. Jia, G. F. Chen, D. M. Wang, J. B. He, T. Qian, J. P. Hu, H. Ding, and S. C. Wang, Phys. Rev. Lett. **109**, 037003(2012).

[21] W. Bao, Q.-Z. Huang, G.-F. Chen, M. A. Green, D.-M. Wang, J.-B. He, and Y.-M. Qiu, Chin. Phys. Lett. **28**, 086104(2011).

[22] W. Li, H. Ding, Z. Li, P. Deng, K. Chang, K. He, S. Ji, L. Wang, X. Ma, J.-P. Hu, X. Chen, and Q.-K. Xue, Phys. Rev. Lett. **109**, 057003(2012).

[23] J. Zhao, H. Cao, E. Bourret-Courchesne, D. H. Lee, and R. J. Birgeneau, Phys. Rev. Lett. **109**, 267003(2012).

[24] X. Ding, D. Fang, Z. Wang, H. Yang, J. Liu, Q. Deng, G. Ma, C. Meng, Y. Hu, and H. H. Wen, Nat. Commun. **4**, 1897(2013).

[25] Z. P. Yin, K. Haule, and G. Kotliar, Nat. Mater. **10**, 932(2011).

[26] P. A. Lee, N. Nagaosa, and X. G. Wen, Rev. Mod. Phys. **78**, 17(2006).

[27] W. Li, H. Ding, P. Deng, K. Chang, C. Song, K. He, L. Wang, X. Ma, J.-P. Hu, X. Chen, and Q.-K. Xue, Nat. Phys. **8**, 126(2012).

[28] Z.-W. Wang, Z. Wang, Y.-J. Song, C. Ma, Y. Cai, Z. Chen, H.-F. Tian, H.-X. Yang, G.-F. Chen, and J.-Q. Li, J. Phys. Chem. C **116**, 17847(2012).

[29] G. R. Stewart, Rev. Mod. Phys. **83**, 1589(2011).

[30] D. P. Shoemaker, D. Y. Chung, H. Claus, M. C. Francisco, S. Avci, A. Llobet, and M. G. Kanatzidis, Phys. Rev. B **86**, 184511(2012).

[31] J. Guo, X.-J. Chen, J. Dai, C. Zhang, J. Guo, X. Chen, Q. Wu, D. Gu, P. Gao, L. Yang, K. Yang, X. Dai, H.-K. Mao, L. Sun, and Z. Zhao, Phys. Rev. Lett. **108**, 197001(2012).

[32] J. Demsar, R. D. Averitt, K. H. Ahn, M. J. Graf, S. A. Trugman, V. V. Kabanov, J. L. Sarrao, and A. J. Taylor, Phys. Rev. Lett. **91**, 027401(2003).

[33] Y. C. Wen, K. J. Wang, H. H. Chang, J. Y. Luo, C. C. Shen, H. L. Liu, C. K. Sun, M. J. Wang, and M. K. Wu, Phys. Rev. Lett. **108**, 267002(2012).

[34] D. H. Torchinsky, G. F. Chen, J. L. Luo, N. L. Wang, and N. Gedik, Phys. Rev. Lett. **105**, 027005(2010).

[35] D. H. Torchinsky, J. W. McIver, D. Hsieh, G. F. Chen, J. L. Luo, N. L. Wang, and N. Gedik, Phys. Rev. B **84**, 104518(2011).

[36] T. Mertelj, V. V. Kabanov, C. Gadermaier, N. D. Zhigadlo, S. Katrych, J. Karpinski, and D. Mihailovic, Phys. Rev. Lett. **102**, 117002(2009).

[37] T. Mertelj, P. Kusar, V. V. Kabanov, L. Stojchevska, N. D. Zhigadlo, S. Katrych, Z. Bukowski, J. Karpinski, S. Weyeneth, and D. Mihailovic, Phys. Rev. B **81**, 224504(2010).

[38] S. Dal Conte, C. Giannetti, G. Coslovich, F. Cilento, D. Bossini, T. Abebaw, F. Banfi, G. Ferrini, H. Eisaki, M. Greven, A. Damascelli, D. van der Marel, and F. Parmigiani, Science **335**, 1600(2012).

[39] P. Kusar, V. V. Kabanov, J. Demsar, T. Mertelj, S. Sugai, and D. Mihailovic, Phys. Rev. Lett. **101**, 227001(2008).

[40] E. E. M. Chia, D. Talbayev, J.-X. Zhu, H. Q. Yuan, T. Park, J. D. Thompson, C. Panagopoulos, G. F. Chen, J. L. Luo, N. L. Wang, and A. J. Taylor, Phys. Rev. Lett. **104**, 027003(2010).

[41] K. W. Kim, A. Pashkin, H. Schaefer, M. Beyer, M. Porer, T. Wolf, C. Bernhard, J. Demsar, R. Huber, and A. Leitenstorfer, Nat. Mater. **11**, 497(2012).

[42] C. W. Luo, I. H. Wu, P. C. Cheng, J. Y. Lin, K. H. Wu, T. M. Uen, J. Y. Juang, T. Kobayashi, D. A. Chareev, O. S. Volkova, and A. N. Vasiliev, Phys. Rev. Lett. **108**, 257006(2012).

[43] C. W. Luo, I. H. Wu, P. C. Cheng, J. Y. Lin, K. H. Wu, T. M. Uen, J. Y. Juang, T. Kobayashi, Y. C. Wen, T. W. Huang, K. W. Yeh, M. K. Wu, D. A. Chareev, O. S. Volkova, and A. N. Vasiliev, New J. Phys. **14**, 103053(2012).

[44] C. Zhang, W. Li, B. Gray, B. He, Y. Wang, F. Yang, X. Wang, J. Chakhalian, and M. Xiao, J. Appl. Phys. **113**, 083901(2013).

[45] V. V. Kabanov, J. Demsar, B. Podobnik, and D. Mihailovic, Phys. Rev. B **59**, 1497(1999).

[46] A. Rothwarf, and B. N. Taylor, Phys. Rev. Lett. **19**, 27(1967).

[47] G. Coslovich, C. Giannetti, F. Cilento, S. Dal Conte, G. Ferrini, P. Galinetto, M. Greven, H. Eisaki, M. Raichle, R. Liang, A. Damascelli, and F. Parmigiani, Phys. Rev. B **83**, 064519(2011).

[48] R. M. Fernandes, L. H. VanBebber, S. Bhattacharya, P. Chandra, V. Keppens, D. Mandrus, M. A. McGuire, B. C. Sales, A. S. Sefat, and J. Schmalian, Phys. Rev. Lett. **105**, 157003(2010).

[49] D. Hsieh, F. Mahmood, D. H. Torchinsky, G. Cao, and N. Gedik, Phys. Rev. B **86**, 035128(2012).

[50] A. Ignatov, A. Kumar, P. Lubik, R. H. Yuan, W. T. Guo, N. L. Wang, K. Rabe, and G. Blumberg, Phys. Rev. B **86**, 134107(2012).

[51] H.-T. Kim, Y. W. Lee, B.-J. Kim, B.-G. Chae, S. J. Yun, K.-Y. Kang, K.-J. Han, K.-J. Yee, and Y.-S. Lim, Phys. Rev. Lett. **97**, 266401(2006).

[52] S. Jagdeep, *Ultrafast Spectroscopy of Semiconductors and Semiconductor Nanostructures* (Springer-Verlag, Berlin, 1996), 2nd edn.

**Figure 1**

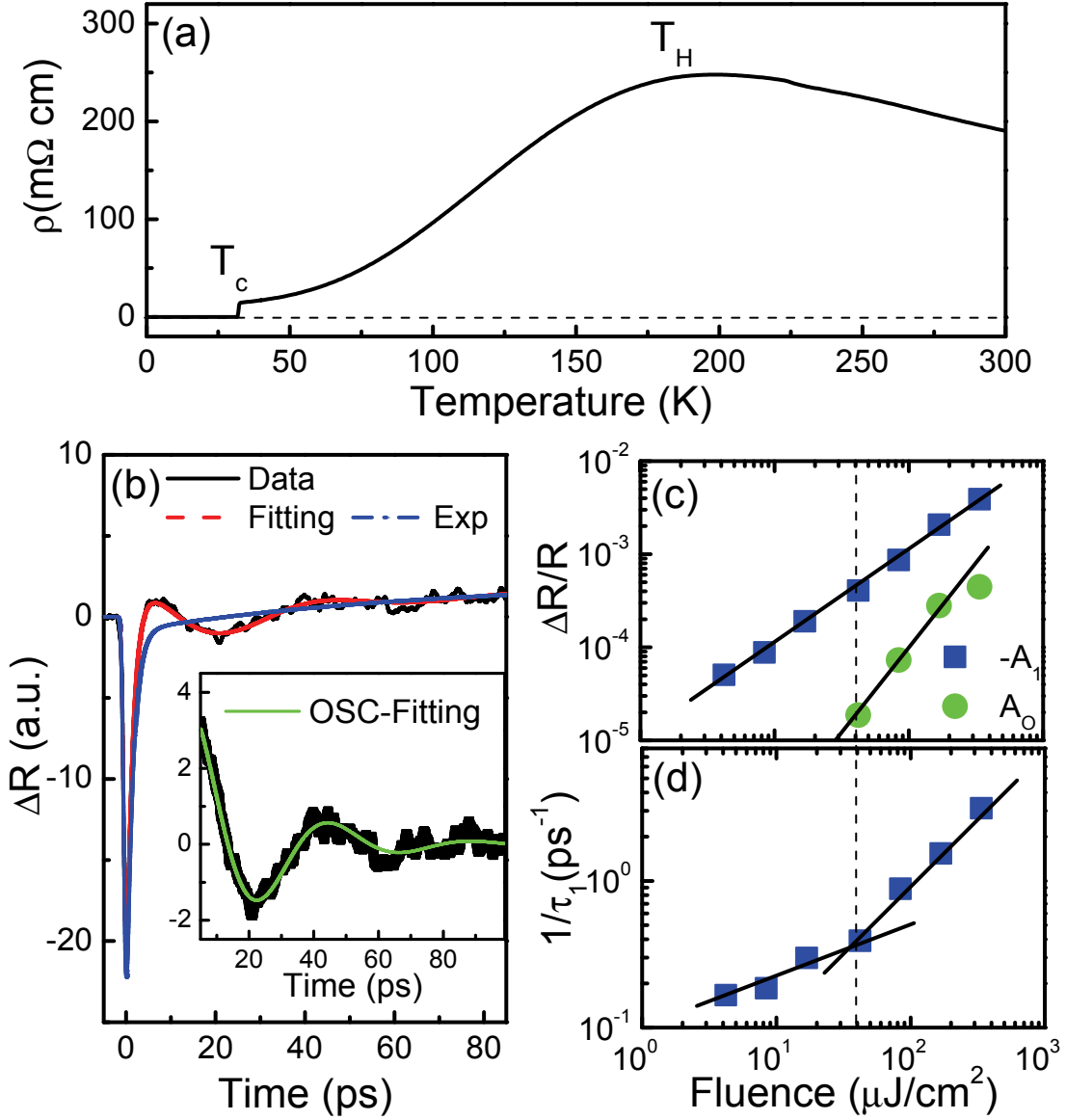

FIG. 1 (Color online) Power-dependent QP dynamics at superconducting phase. (a) Temperature-dependent resistivity of a superconducting $K_xFe_{2-y}Se_2$ single crystal. Besides a superconducting transition at $T_c \sim 32$ K, a resistivity hump appears at $T_H \sim 180$ K. (b) Time-resolved reflectivity change recorded at 5 K with excitation flux of ~ 200 µJ/cm². The decay curve is analyzed with Eq.1. The exponential decay component and the oscillatory component (inset) are highlighted. (c) The fluence-dependent amplitudes of the dominant fast decay component ($A_1$) and oscillatory component ($A_O$). (d) The decay rate of the fast component is plotted as a function of the incident fluence.

**Figure 2**

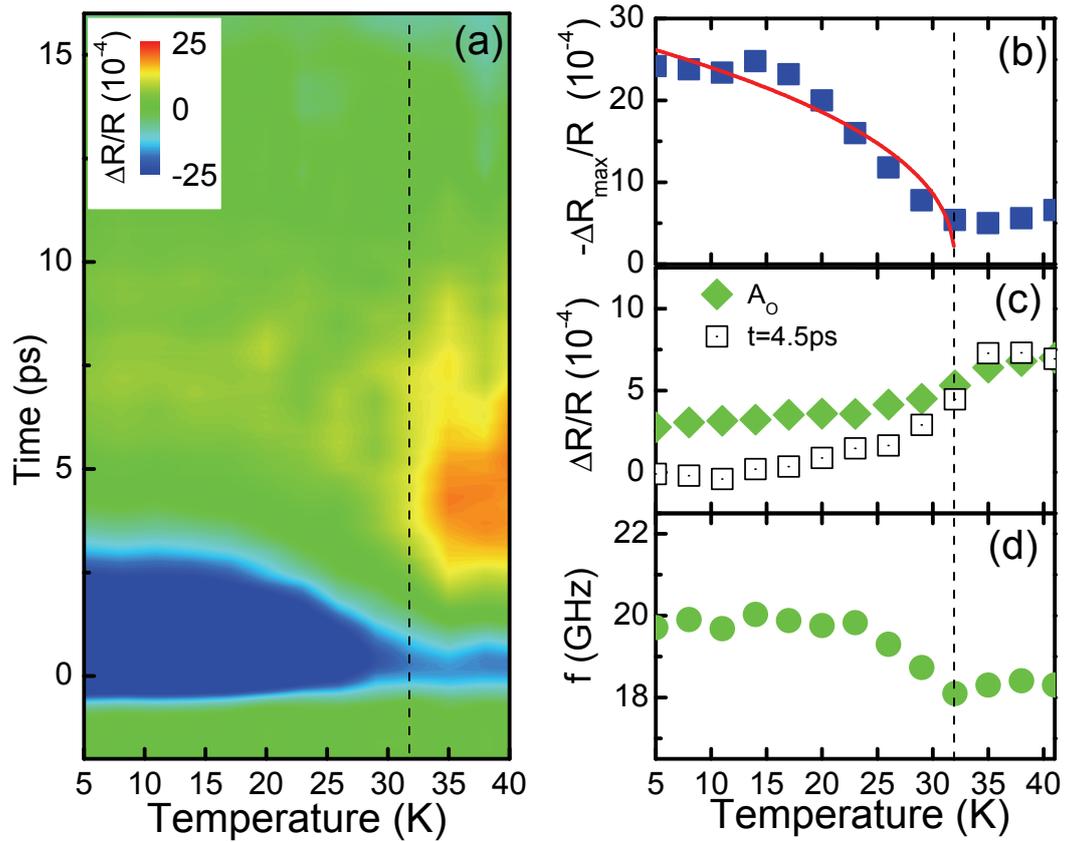

FIG. 2 (Color online) Temperature-dependent QP dynamics in KFS superconductors. (a) The differential reflectivity change is plotted as a function of temperature and delay time. The maximum value of the initial reflectivity change (b), the signal magnitude at t=4.5 ps and the fitted amplitude of oscillatory component (c), and the oscillatory frequency (d) are plotted versus temperature. The dashed lines indicate the superconducting transition temperature.

**Figure 3**

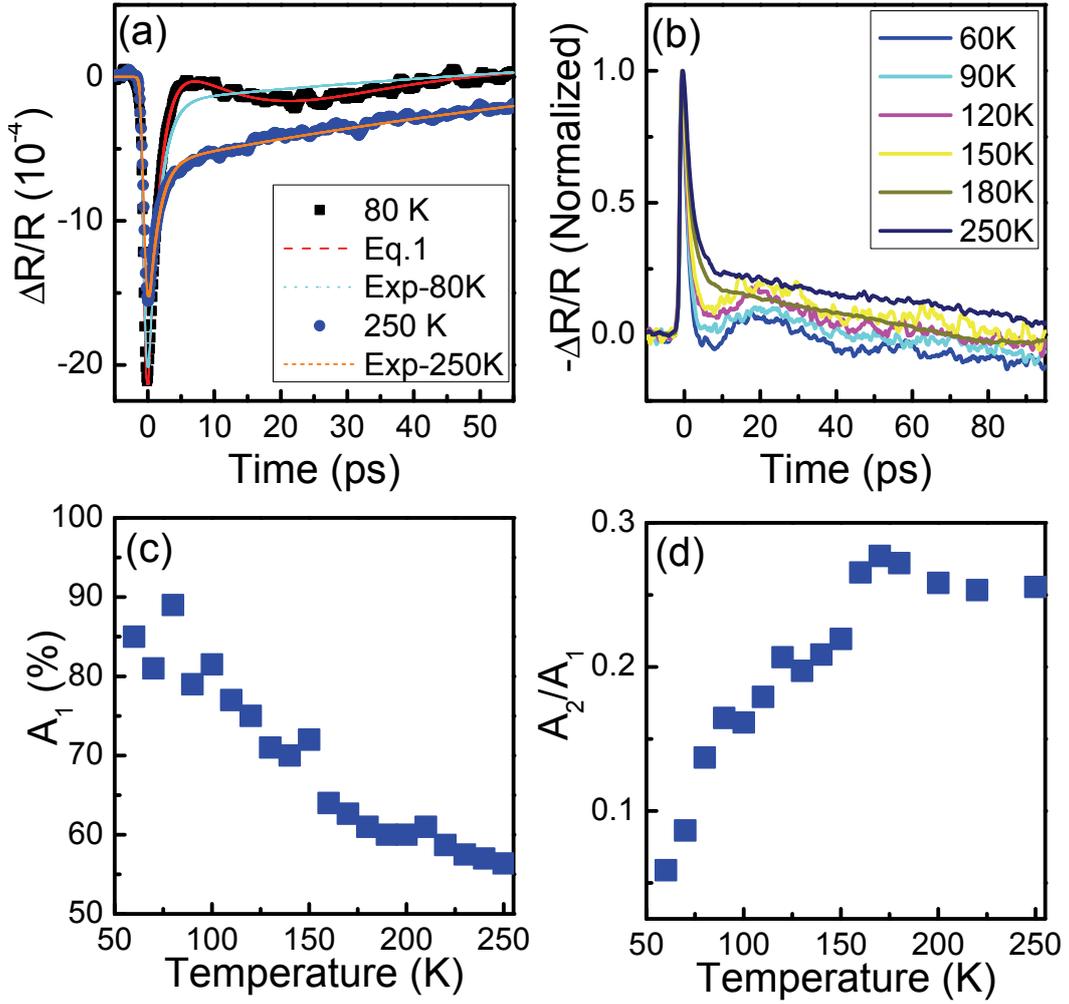

FIG. 3 (Color online) QD dynamics at the normal state. (a) The time-resolved differential reflectivity recorded at 80 K and 250 K, respectively. (b) The normalized time-resolved differential reflectivity change recorded at different temperatures. (c) The temperature-dependent amplitude ratio of the fast decay component. (d) The temperature-dependent A2/A1.

**Figure 4**

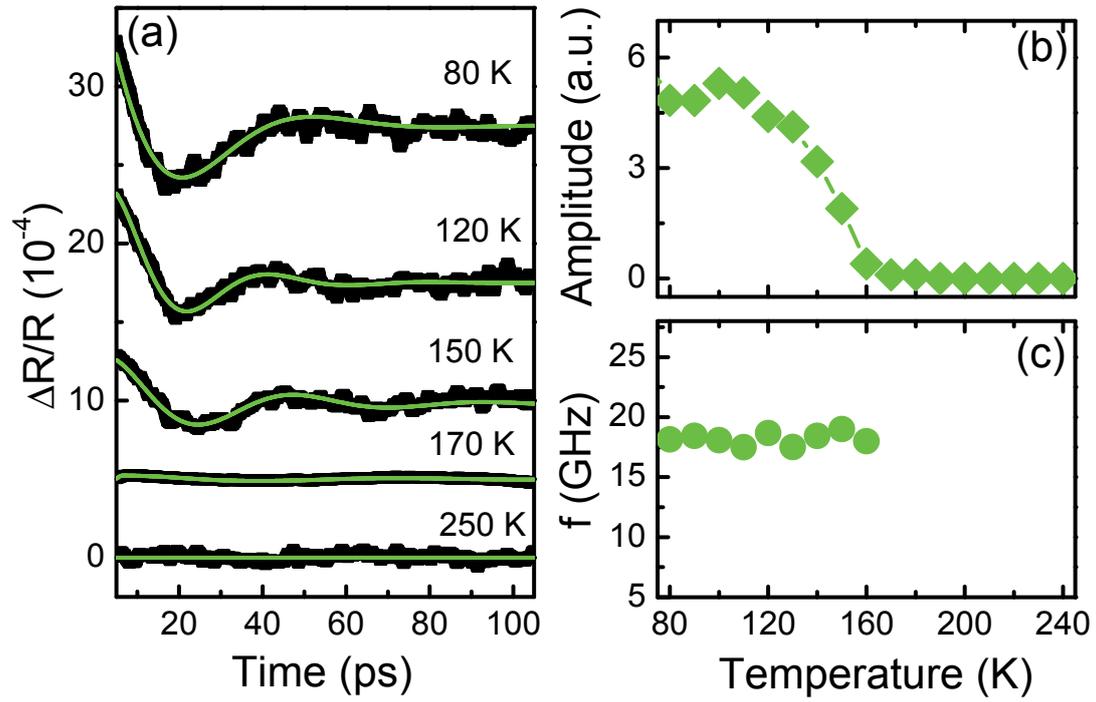

FIG. 4 (Color online) Temperature dependence of the oscillatory component. (a) The oscillatory components after subtraction of the exponential decay components recorded at different temperatures. The traces are vertically shifted for clarity. Temperature-dependent amplitude (b) and frequency (c) of the oscillatory component.

# Orbital-Selective Mottness in $K_xFe_{2-y}Se_2$ Superconductors Revealed by Pump-Probe Spectroscopy

(Supplementary Material)

**(1) Experimental details**

The samples used in this study were prepared by a rapid quenching approach as reported previously [1, 2]. Great care was taken during sample loading to avoid exposing the samples to the ambient environment. We put the optical cryostat into the argon tank where the samples were stored. The cleaved samples were glued onto the cold finger and the cryostat was sealed before it was taken outside. After taking outside, the cryostat was then vacuumed immediately for temperature-dependent optical studies.

The time-resolved traces of $\Delta R/R$ were recorded with the dual-color pump-probe measurements. Low repetition (1 kHz) ultrafast pulses from a Ti:sapphire regenerative amplifier (Libra, Coherent; pulse duration ~ 100 fs) were employed to perform the optical studies. $\Delta R/R$ was monitored with the probe photon energy of 1.55 eV, while the second-harmonic beam with a photon energy of 3.1 eV was used to excite the sample in this study. The pump beam (~1 mm in diameter) and the probe beam (~0.2 mm in diameter) coincide on the sample. The fluence of the probe beam was set to be less than 1/20 of that of the pump beam. The transient optical data were acquired with a balanced detector and a lock-in amplifier. Bandpass filters were employed to exclude the scattered pump light. With such configuration, the signal-to-noise ratio for detecting $\Delta R/R$ has reached a level of better than $10^{-5}$. We have carefully done cycle tests to exclude the aging effect. The experimental data can be well reproduced within 48 hours after sample loading. We have carried out the experimental measurements on over 10 samples from 4 batches. The major features are almost the same from batch to batch except slight differences on the amplitudes of the components and the exact temperature where the oscillatory component disappears.

**(2) Quasiparticle dynamics under weak perturbation**

The dynamics of photo-excited quasiparticles (QPs) in superconductors has been widely studied with excitation in weak perturbation regime. The relaxation of photo-excited superconductors can be well explained by the Rothwarf-Taylor (RT) model [3-5]. Figure S1 shows the data obtained from the KFS superconductors under weak excitation (~ 12 μJ /cm$^2$). The dynamics can be approximately analyzed by an exponential decay function as $\Delta R(t)/R = A_1 \exp^{-t/\tau_1} + C$ with a constant C for the sub-ns background due to thermal equilibrium. In the RT model [3-5], the temperature-dependent amplitude $A_1$ can be evaluated as

$$[A_1(T)/A(T->0)]^{-1} - 1 \propto n_T \propto \sqrt{\Delta(T)T} \exp(-\Delta(T)/T), \quad (S1)$$

where $T$ is the temperature and $n_T$ is the density of QPs. As shown in Fig. S1 (c), the temperature-dependent amplitude can be well fitted with a superconducting gap value of 12 meV by assuming $\Delta(T) = \Delta_0[1 - T/T_C]^{1/2}$. The good fit indicates that the QP relaxation dynamics in such KFS superconductor samples can be described by the presence of a gap at the Fermi level. The superconducting gap gives rise a bottleneck for the carrier relaxation. A high energy boson with energy $>2\Delta$ is created by two QPs with their energies higher than $\Delta$, which breaks additional Cooper pairs and slows down the QP recombination. With temperature approaching $T_C$, more high energy Bosons are available to regenerate QPs, so that the relaxation lifetime becomes much slower as shown in Fig. 2(d).

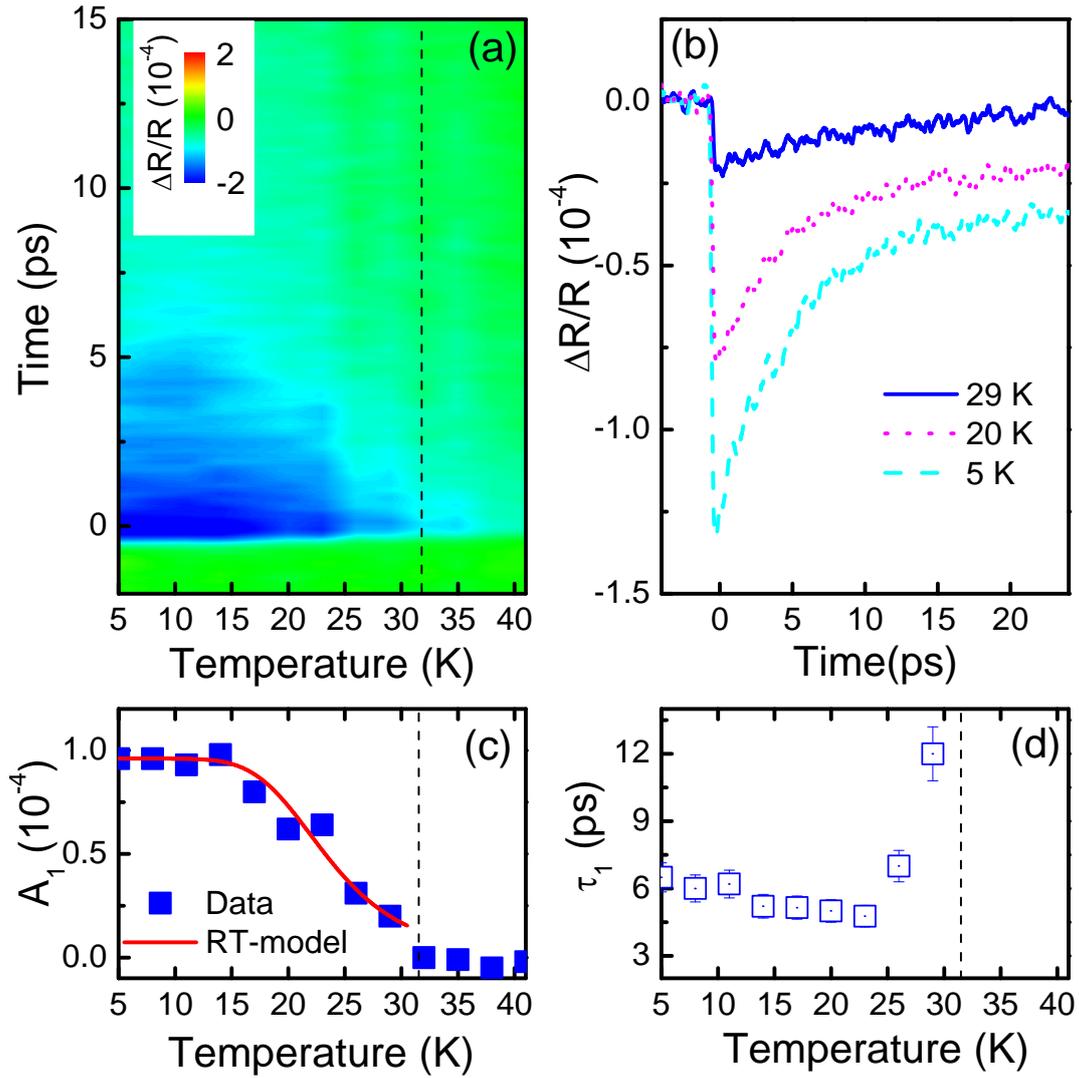

FIG.S1, The QP dynamics under weak perturbation. (a) $\Delta R/R$ signal is plotted as functions of temperature and delay time. (b) The time-resolved $\Delta R/R$ curves recorded at different temperatures subtracted by the signal recorded at 32 K. The maximum signal amplitude (c) and the decay lifetime (d) are plotted versus the temperature. The solid line in (c) is the fitting curve with the RT-model.

**(3) Fluence-dependent QP dynamics**

The time-resolved traces of $\Delta R/R$ are recorded with different exciton fluences. As shown in Fig. S2, the amplitudes of both the exponential decay and oscillatory component are dependent on the exciton fluence. Fitting the raw data to Eq.1 gives the fluence dependence as shown in Figs. 1(c) & 1(d) in the main text. As we have discussed in the main text, the fluence dependence of oscillatory component shows a threshold behavior. Moreover, the decay lifetime is also strongly dependent on the incident fluence. The significant increase of the decay rate with the increasing excitation fluence indicates an important role played by the electron-electron interactions for the dynamics of photo-excited QPs in superconducting KFS samples[6].

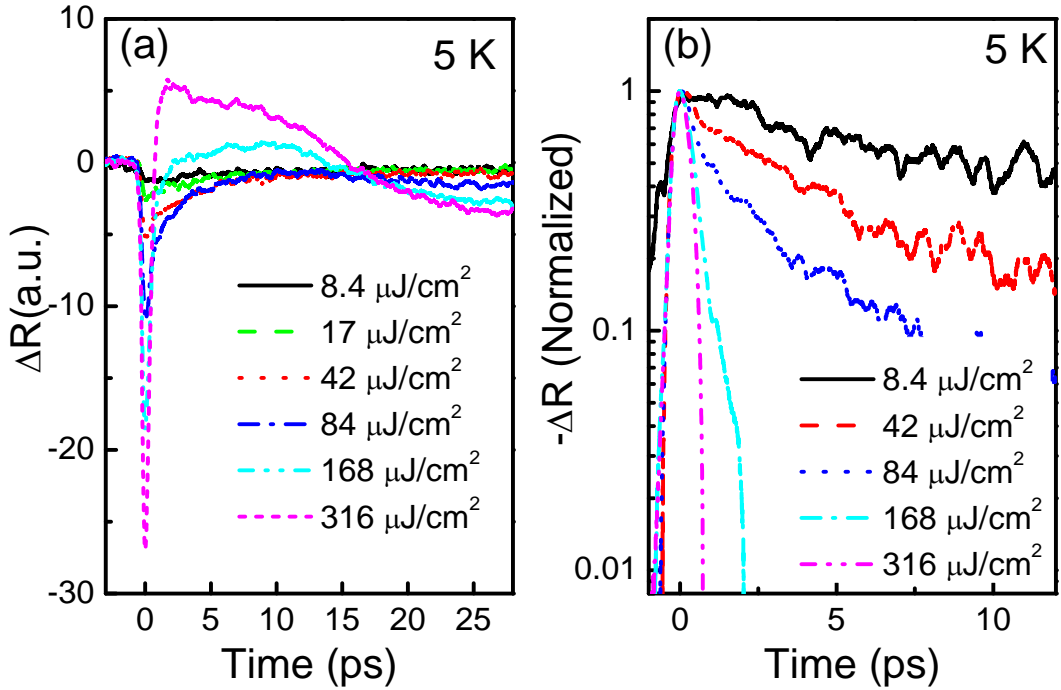

FIG.S2, (a) Time-resolved $\Delta R/R$ signals of the KFS superconductors recorded at 5 K with different excitation fluences; (b) Normalized traces to highlight the significant fluence-dependent fast decay component.

**(4) Fourier transform of the oscillatory components**

To quantify the coherent-phonon-induced oscillatory components in the time-resolved traces of $\Delta R/R$, we subtracted the exponential decay components from the total signals and analyze the residual part with a Fourier transformation. As shown in Fig. S3, the amplitude has clear temperature–dependence, while the oscillation frequency is basically independent of the temperature.

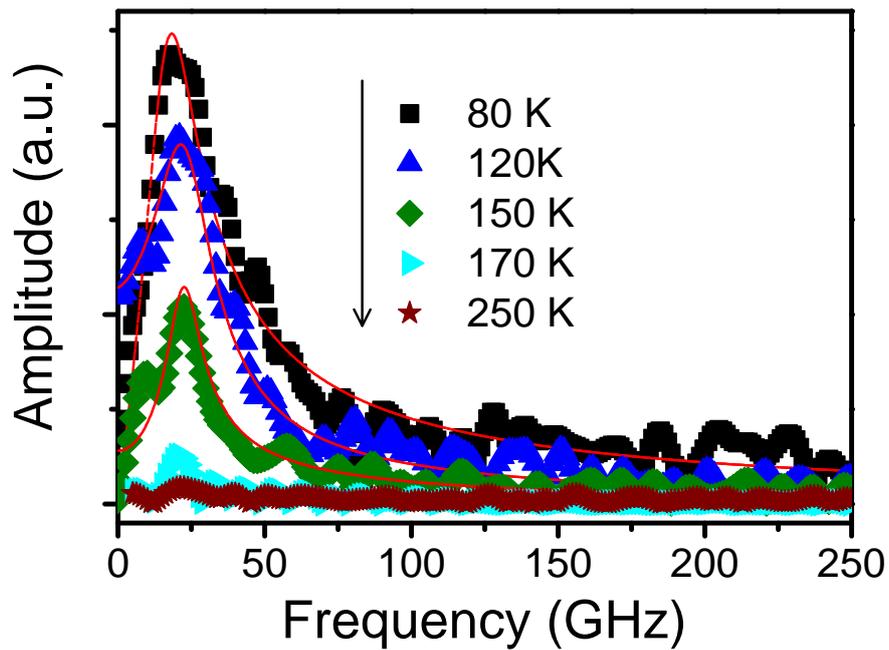

FIG.S3, Fourier transformation of the oscillatory components obtained at different temperatures.

# References

[1] X. Ding, D. Fang, Z. Wang, H. Yang, J. Liu, Q. Deng, G. Ma, C. Meng, Y. Hu, and H. H. Wen, Nat. Commun. **4**, 1897(2013).

[2] H.-H. Wen, Rep. Prog. Phys. **75**, 112501(2012).

[3] A. Rothwarf, and B. N. Taylor, Phys. Rev. Lett. **19**, 27(1967).

[4] E. E. M. Chia, D. Talbayev, J.-X. Zhu, H. Q. Yuan, T. Park, J. D. Thompson, C. Panagopoulos, G. F. Chen, J. L. Luo, N. L. Wang, and A. J. Taylor, Phys. Rev. Lett. **104**, 027003(2010).

[5] G. Coslovich, C. Giannetti, F. Cilento, S. Dal Conte, G. Ferrini, P. Galinetto, M. Greven, H. Eisaki, M. Raichle, R. Liang, A. Damascelli, and F. Parmigiani, Phys. Rev. B **83**, 064519(2011).

[6] D. H. Torchinsky, G. F. Chen, J. L. Luo, N. L. Wang, and N. Gedik, Phys. Rev. Lett. **105**, 027005(2010).